\documentclass[12pt,thmsa]{article}

\input tcilatex
\QQQ{Language}{
American English
}

\begin{document}

\begin{center}
The Theory of vacuum and some practical results.

A.V.Rykov

Chief of Seismometry lab. of IPE RAN, Moscow, Russia.
\end{center}

\textit{Vacuum, where matter exists is an objective reality of Nature. It
has a structure consists of electrical massless dipoles. This structure is
responsible for gravitation, inertia and propagation of light . The
structure can be influenced by the electrical, magnetic forces and by
radiation and thas control the gravitation and inertia.}

\textit{\ }Void is only void and nothing more. Void cannot have any physical
properties. For example, vacuum has physical parameters, i.e. dielectric and
magnetic penetrability. That's why vacuum cannot be a void or empty space of
the Universe. Let's consider the structure vacuum in details. At first we'll
remove a blunder of physics presented by Coulomb's formula. It lies in the
fact that permittivity of vacuum were put to the denominator of formulas for
electric and magnetic forces. We'l introduce inverse values:

$\eta =\frac 1\mu =1.0000000028\cdot 10^7[a^2kg^{-1}m^{-1}s^2].$ It is a
magnetic constant of vacuum equal to inverse value of magnetic permeability. 
$\xi =\frac 1\varepsilon =8.98755179\cdot 10^9[a^{-2}m^3kg\cdot s^{-4}]$ is
a dielectric constant of vacuum equal to inverse value of dielectric
permittivity. Newton's and Coulomb's formulas get an identical view. Speed
of light gets more logical idea $c=\sqrt{\eta \xi }$ .

Experimental physics presents necessary data for the study of vacuum. We
mean the data on photoeffects in vacuum, on nuclei and nucleons [Karjakin
N.I. and others, 1964]. Let's remind the values of gamma-quanta energies: 1,
137, 1836, 3672 MeV ($2m_ec^2,137\cdot 2m_ec^2,1836\cdot 2m_ec^2,1836\cdot
4m_ec^2$). This series of energy gives a valuable information for the
physical ideas about the structure of vacuum and matter [Rykov A.V., 2001].

Gamma-quanta of $\nu $ frequency deforms the structure of cosmic vacuum.
Being within the size of $r_e$ between its elements, gamma-quanta creates a
deformation $\Delta r_e$ . The deformation energy will be $e_oE\Delta r_e$ $%
, $ where $e_o$ is a elementary charge, $E$ - is electrical intensity of the
structure. Equation of the energy will be

\begin{center}
$h\nu =e_oE\Delta r_e$ (1),
\end{center}

where $h$ - is a Plank's constant. Deformation is function of time

\begin{center}
$\Delta r_e=\Delta [r_e\sin (2\pi \nu t)]=2\pi \nu r_e\Delta t\cos (2\pi \nu
t)$ (2).
\end{center}

Let's define the intensity of electrical field, where N is some coefficient
of proportionality:

\begin{center}
$E=N\xi \frac{e_o}{r_e^2}$ (3).
\end{center}

Let's put the obtained expressions, amplitude from (2) and intensity from
(3) to (1):

\begin{center}
$h=2\pi Ne_o^2\xi \frac 1{r_e/\Delta t}$ (4).
\end{center}

We can assume quite naturally that $r_e/\Delta t=c$ - is speed of light.
Let's find an unknown quantity:

\begin{center}
$N=\frac h{2\pi e_o^2r_q}=137.035990905=\alpha ^{-1}$ ! (5),
\end{center}

where $r_q=\sqrt{\xi /\eta }$ . We have got a well known formula of Plank's
constant:

\begin{center}
$h=2\pi e_o^2\alpha ^{-1}\sqrt{\xi /\eta }=6.6260755(40)\cdot 10^{-34}$ (6).
\end{center}

On this stage we should clear a situation with chose of numerical values for 
$h$ or $\alpha ^{-1}$ . All next values are calculated on the base of $h$.
But the $\alpha ^{-1}$ is in reality more fundamental then $h$, because the
last one is derivative from $e_o^{},\alpha ^{-1},\xi $ $,\eta $ - vacuum
parameters. The choice made here is based upon this quite new study of
vacuum.

Gamma--quantum of energy $w\geq 1$ MeV interacting with vacuum changes a
''virtual'' electron-positron pair to the real ones. The energy equation of
this change is:

\begin{center}
$w=h\nu _{rb}=\xi \frac{e_o^2}{r_e}$ (7),$^{}$
\end{center}

where $r_e$ - distance between charges (+) and (-) of vacuum structure, $\nu
_{rb}=2.4892126289\cdot 10^{20}$ Hz - ''red border'' for frequency of
gamma-quantum . The last exact value is determined below. Let's find $r_e$ :

\begin{center}
$r_e=\frac{\xi \alpha }{2\pi r_q\nu _{rb}}=\frac{c\alpha }{2\pi \nu _{rb}}%
=1.398763188\cdot 10^{-15}m$ (8).
\end{center}

We have from (2) $\Delta r_e=2\pi \nu _{rb}r_e\Delta t=\frac{2\pi \nu
_{rb}r_e}cr_e=\alpha \cdot r_e$ under assumption $r_e/\Delta t=c$. In other
words, it is the limit of the vacuum deformation above what a rupture of
structure ties occurred:

\begin{center}
$\Delta r_e=\alpha \cdot r_e=1.020726874\cdot 10^{-17}m$ (9).
\end{center}

The exact value for $\nu _{rb}=\frac c{2\pi r_e\alpha ^{-1}}=2.48921263\cdot
10^{20}Hz$ . Deformation of structure lower than the given value has
electroelastic character. Let's find the coefficient of elasticity $b$ from
a forth equation:

\begin{center}
$f=b\Delta r_{rb}=\xi \frac{e_o^2}{r_e^2}$ $,$ $b=1.155219829\cdot
10^{19}[kg\cdot s^{-2}]$ (10).
\end{center}

Another useful parameters of vacuum will be next:

\begin{center}
$E_\sigma =\sqrt{\gamma \xi }=0.77440463$ $[a^{-1}m^3s^{-3}]$ (11) and

$S=\alpha ^{-2}\frac{e_o}{4\pi r_e^4}=6.254509137\cdot 10^{43}[Q\cdot m^{-4}$
(12).
\end{center}

The names for this parameters are not yet known.

To that stage we get the main parameters of the vacuum structure.

Some consequences from the vacuum structure.

1. Dielectric vacuum media has a tied charges. The moving charge generate a
Maxwell's displacement current $j$. This current generate magnetic strength $%
\overline{dH}=\frac 1c\overline{j}\stackrel{}{}$ where $\overline{j}=\frac
1{4\pi }\frac{d\overline{E}}{dt}$. The $\overline{H}$ is necessary magnetic
component to the $\overline{E}$ for the Electromagnetic wave (light). The
vacuum structure is natural media for light excitation and propagation in
space.

2. Nature of quantum mechanic's are defined by the vacuum structure.
Compton's length of an electron is equal $\lambda =\frac h{m_oc}=4\pi
(r_e+\Delta r_e)\cdot \alpha ^{-1}=2.42631057573\cdot 10^{-12}m$ (13).

The expression - $4\pi (r_e+\Delta r_e)\cdot \alpha ^{-1}=2.42631057573\cdot
10^{-12}m$ - completely defined by parameters of vacuum. Another words - the
permitted electron orbits in atoms are defined by structure of vacuum (the
nature of quantum mechanic).

3. De Brogl's wave $\lambda =h/mV$ $.$ Plank's constant is completely
defined by the parameters of vacuum - formula (6). This leads to the sin-way
of a particle trajctory what confine the diffraction appearance in a nature.

4. Electron mass can be produced by exited vacuum $m_e=\frac{e_o^2}{2\eta
(r_e+\Delta r_e)}=9.1093897427\cdot 10^{-31}kg$ .

5. The laws of Newton and Coulomb can be united next way.

$f=\gamma \frac{m^2}{R^2}=\xi \frac{q^2}{R^2}$ and $\rho =\sqrt{\frac \gamma
\xi }=8.6164135164\cdot 10^{-11}[Q\cdot kg^{-1}]$- the electrical charge of
one kg of any mass. The same value may be presented thorough a micro
parameters - $\rho =e_o\sqrt{\frac{2\pi \gamma }{ch\alpha }}=8.6164135\cdot
10^{-11}$ $.$

6. Gravitational constant is defined by parameters of vacuum $\gamma =\frac{%
\xi e_o^2}{m_x}=6.67259049725\cdot 10^{-11}$[kg$^{-1}$m$^3$s$^{-2}$] where $%
m_x=m_{Pl}\sqrt{\alpha }=1.8594480544\cdot 10^{-9}$kg, $m_{Pl}$ - Plank
mass. It is indirect evidence of electrical nature of gravitation. The
vacuum has a very smal superiority one charge respect to other. Correctly in
21 sign of electron charge. On the law of Faraday induction a charged media
attracts all bodis to each other. Ratio of mass attraction and vacuum
Coulomb repultion in Universe forms $\Lambda $ - coefficient in Einstein's
theory.

7. Acceleration of mass or of gravity force creates a vacuum deformation and
the last one can be calculated by formula:

\begin{center}
$\Delta r_{a,g}=\sqrt{\frac{a,g}{4\pi E_\sigma S}}$ (13)
\end{center}

For instance the deformation under Earth gravity would be $\Delta
r_{Earth}=1.2703\cdot 10^{-22}m.$ A force of accelerated mass is determined
by $f=am=b\cdot \Delta r_a$ and is an elastic force of resistance to
accelerated motion.

8. Maximum of a gravity acceleration is defined from $g_{\max }=4\pi
E_\sigma S\cdot (\Delta r_{rb})^2=6.3414723\cdot 10^{10}m\cdot s^{-2}$ . It
defines ''horizons of events'' and evaporation of ''Black Holes'' discovered
theoretically by Hawking (birth of electrons and positrons from vacuum).

9. Indirect evidence of reality of all represented hire consideration we get
from next correlations:

\begin{center}
$b\cdot \Delta r_{rb}=m_xg_{\max }$ (14),

$m_x=1.859480544\cdot 10^{-9}$ kg.
\end{center}

We are already met this mass at point (6). What does it mean? First of all
this mass can be a smallest ''black hole'' with size $r_x=\sqrt{\gamma \frac{%
m_x}{g_{\max }}}=1.39876319\cdot 10^{-15}$ m. Secondly we find a remarkable
coincidence $\rho m_x=1.60220\cdot 10^{-19}$ Q - very close to the electron
charge. Again we get indirect evidence in behalf of the represented vacuum
paradigm. All values like $\rho ,e_o,m_x,\alpha ,m_{Pl}$ appears to be very
close tightened each with other.

10. And now there is the most fantastic a practical outcome: as we see
gravitation and inertia forces are connected with a deformation of vacuum
structure. And becourse of this circumstance we can control those forces,
for instance, by electrical intensity. However, to compensate the Earth
gravity it is necessary to applied $E=1.1402\cdot 10^{10}$ V/m. That is
impossible. The experience conducted by russian scientist Roschin V.V. and
Godin S.M. [Roschin,\ Godin, 2000] shows that alternative magnetic intensity 
$H$ may reduce the gravity and inertia vacuum deformations. Strong magnets
rotated up to velocity 550 rpm. After achieving that velocity the rotor was
speeding rotation without any outer power supply (decrease the momentum of
inertia). It is needed to add power consumption about 6 KWatt from rotating
magnets to brake the accelerated rotation to keep it steady. The reduction
of the vacuum deformation is possible to estimate by formulas. According to
Maxwell's we have $E_z=l\frac{dB_x}{dt}.$ Approximatly $E_z=v\cdot B_x=v\eta
H$ , where $v$ - rotation velocity. Eventually we get $\Delta r_E=\frac{%
e_oE_z}b=\frac{e_o\eta H}bv$ m. Thus we can compensate the gravety
acceleration as $\Delta r_{ng}=\Delta r_{gEarth}-\Delta r_E$ and reduce the
gravity force of Earth.

Conclusion.

It is discovered the structure of vacuum - the necessary media for Universe
to live in. The light (EMW), gravity, inertia, atoms of matter etc. may not
to exist without the media -structured vacuum. The structure of vacuum has
many applications in different fields of science.

\begin{center}
Literature.
\end{center}

\section{Karyakin N.I. and others. Abridged guide on physics.M.:1964, 560
p.(in russian).}

\section{Roschin V.V., Godin S.M. Experimental research of physical effects
in dynamic magnetic system. The letters to MTPh, St.Pb, 2000, v.26, is.24,
73-81 p.(in russian).}

\section{Rykov A.V. Principles of natural physics. OIPhE RAS, M.:2001, 58
p.(in russian).}

\end{document}